\makeatletter\let\ifGm@compatii\relax\makeatother
\documentclass[pre,aps,twocolumn,floatfix,superscriptaddress]{revtex4-2}
\pdfoutput=1

\usepackage{amsfonts,amsmath, color,graphicx, subfigure,amssymb}
\usepackage{eso-pic,calc}
\usepackage{epsfig}
\usepackage{latexsym}
\usepackage{bm}

\def\prl#1#2#3{{ Phys. Rev. Lett.} {\bf #1}, #2 (#3)}

\def\pre#1#2#3{Phys. Rev. E {\bf #1}, #2 (#3)}
\def\prx#1#2#3{Phys. Rev. X {\bf #1}, #2 (#3)}

\def\pra#1#2#3{Phys. Rev. A {\bf #1}, #2 (#3)}

\def\sc#1#2#3{Science {\bf #1}, #2 (#3)}

\begin{document}
\title{Linking space-time correlations for  a class of self-organized critical systems}

\author{Naveen Kumar}
\affiliation{Department of Physics \& Astronomical Sciences, Central University of Jammu, Samba 181 143, India}

\author{Suram Singh}
\affiliation{Department of Physics \& Astronomical Sciences, Central University of Jammu, Samba 181 143, India}

\author{Avinash Chand Yadav}
\affiliation{Department of Physics, Institute of Science,  Banaras Hindu University, Varanasi 221 005, India}

\begin{abstract}
{The hypothesis of self-organized criticality explains the existence of  long-range `space-time' correlations, observed inseparably in many natural dynamical systems. A simple link between these correlations is yet unclear, particularly in fluctuations at `external drive' time scales. As an example, we consider a class of sandpile models displaying non-trivial correlations.  Employing the scaling methods, we demonstrate the computation of spatial correlation by establishing a link between local and global temporal correlations. }
\end{abstract}

\maketitle

\section{Introduction}
The existence of ``long-range'' space-time correlations is a generic behavior observed inseparably in many natural dynamical systems far from equilibrium. Striking examples have been found in diverse contexts, spanning from propagation of spatio-temporal neuronal activity~\cite{Siwy_2002, Lee_2005, Pettersen_2014, Das_2019} to height fluctuations in the interface depinning pulled at boundary~\cite{Maslov_1996} and the release of seismic activity in earthquakes~\cite{Burridge_1967, Vieira_1992, Vieira_2000, Christensen_1992, Danon_2002, Davidsen_2002}. One widely-applicable mechanism is the hypothesis of self-organized criticality (SOC)~\cite{Bak_1987, Bak_1996}, which can explain the origin of such features. The SOC refers that a class of non-equilibrium systems respond nonlinearly in the form of critical avalanches when driven slowly. The response exhibits scaling in the form of power-law distribution for the avalanche sizes as well $1/f^{\alpha}$ type power spectral density~\cite{Dutta_1981, Weissman_1988, Milotti, Miguel_2014, Yadav_2013, Yadav_2017, Erland_2011, Balandin_2013, Sposini_2020}. Note that the emergence of such a critical state is spontaneous, caused by self-organization.

Over the past three decades, the SOC has triggered an enormous amount of research. 
However, a simple link is yet not clearly understood for the long-range space and time correlations. Often, an exact analytical treatment of such dynamical systems remains challenging. 
Several studies have separately focused on only one aspect, either critical avalanches or $1/f$ noise; thus lacking a possible link. In some cases, the fluctuations in the avalanche activity~\cite{Laurson_2005}, fluctuations at a much shorter (fast) time scale have been shown to exhibit~$1/f^{\alpha}$ noise. Here,  the spectral exponent can be related to the critical avalanche characteristics, but the signal does not capture fluctuations at a longer or external drive time scale.

In many cases, those displaying critical avalanches, the fluctuations at external drive time scale show trivial $1/f^2$ behavior. A class of SOC models indeed show $1/f$ noise~\cite {Zhang_1999, Maslov_1999, Yadav_2012}, but no long-range spatial correlations; the avalanches are not critical. Although a one-dimensional variant of the sandpile model shows non-trivial correlations~\cite{Davidsen_2002}, the link obtained in terms of a scaling relation lacks a satisfactory analysis. We emphasize that the fluctuations at a longer (slow) time scale has been of particular interest.  For instance, the earthquakes show critical avalanche statistics (the Gutenberg-Richter law) as well as $1/f^{\alpha}$ noise~\cite{Davidsen_2002}. The long-range temporal fluctuations have been observed for vastly longer time scales from days to years, while the instantaneous events occur at a time scale in the order of seconds.

The purpose of this paper is to reveal a simple link for the long-range space-time correlations observed in the fluctuations recorded at the external drive time scale in a class of SOC models. The systems that we examine show  critical avalanches, exhibiting non-trivial temporal correlations. In particular, we show intensive analysis for a  SOC model, earlier studied by Davidsen and Paczuski~\cite{Davidsen_2002}. Our main tool of analysis is the scaling method~\cite{Christensen_2005, Djordje_2011_pre, Manchanda_2013, Yadav_2021} that allows us to understand the spatial correlations by establishing a relationship for temporal correlations between microscopic and macroscopic fluctuations.
The paper structure is as follows. Section~\ref{sec_2} begins with recalling the definition of the model. In Sec.~\ref{sec_3}, we present analytical results for the spectral properties derived from the scaling methods. Our numerical studies well support these results. We establish a link between non-trivial space-time correlations in Sec.~\ref{sec_4}. Section~\ref{sec_5} shows the extent of the spectral features for fluctuations in the Oslo sandpile model~\cite{Christensen_1996}. Finally, we summarize our results in Sec.~ \ref{sec_6}.

\section{Model}{\label{sec_2}}
We consider a simple SOC model introduced by M. de Sousa Vieira~\cite{Vieira_2000}. The model is a variant of the Burridge-Knopoff train model~\cite{Burridge_1967}, where a block of spring chains is pulled at one end, and  this explains the stick-slip dynamics of earthquake faults.  
The model definition is as follows: Consider a one-dimensional (1-D) lattice of size $L$. To each site $i$, assign a continuous variable $f_i$ denoting local force. The state is continuous, but the update time is discrete. The system is initialized by keeping all local forces at the same value such that $0 \le f_i < f_o$, where $f_o$ is a threshold. A site $i$ becomes unstable if the local force $f_i$ exceeds the threshold $f_i\ge f_o$. The dynamics consist of two elementary steps: (i) External slow drive at the left boundary
\begin{equation}
f_1 = f_o +\delta f,\nonumber
\end{equation} 
where $0< \delta f \ll f_o$.  (ii) As a result of the external drive, a site  $i$ may become unstable. Then, a relaxation occurs, and the force is locally redistributed in a conservative manner
\begin{eqnarray}
f_i \to f_{i}^{'} = \phi(f_i - f_o),\nonumber \\
f_{i\pm 1} \to f_{i\pm 1} + \frac{\Delta f}{2}, ~~~{\rm with}~~~ \Delta f = f_i -f_{i}^{'},\nonumber
\end{eqnarray}
where $\phi(x)$ is a nonlinear periodic function. Here, a particular choice of the function is sawtooth  
$\phi(x) = 1 - a ~{\rm modulo}~ (x, 1/a)$. Also, the results do not change if we take $\phi(x)$ to be a uniformly distributed random variable in the unit interval.

The local redistribution of forces may further trigger neighbor sites, and that may do the same. Thus relaxation continues until all forces become stable. The event forms an avalanche. When an avalanche is over, the system is again driven. The separation of time scales between slow drive and instantaneous relaxation excludes the occurrence of interacting avalanches. The boundary at both ends are open such that $f_0 = f_{L+1} = 0$.  
In the simulation, we use $f_o = 1$, $\delta f = 0.1$, and $a = 4$. 
Note that the simple model is deterministic, both in the initial condition and update rules (i.e., no embedded randomness). In turn, the same trajectory arises for each realization and it is not suitable for ensemble average. So we initialize the system by randomly assigning a different constant value to all local forces. We record the fluctuations in the stationary state after discarding transients. We can shorten the transients by initializing the system configuration with a value close to 1.

Our interest is in a quantity that fluctuates at an external drive time scale. Such a relevant microscopic quantity is the fluctuations in the local force as a function of space and time $\xi(x,t) = f_i(t)$. Let us define fluctuations about mean as $\delta \xi = \xi-\langle\xi\rangle$. Then, the two-time auto-correlation function is expressed as
\begin{equation}
\mathcal{C}_{\xi}(x, \tau) = \langle\delta \xi(x,t_1)\delta\xi(x,t_2)\rangle, 
\label{l_corr}
\end{equation} 
where the angular bracket  $\langle \cdot \rangle$ denotes ensemble average. As  $\xi(x,t) \in [0, 1)$, the process is stationary. The time translational symmetry of $\xi$ implies that  $\tau = |t_1-t_2|$.

In addition, a more interesting macroscopic quantity is the total force
$\eta(t) = \sum_{x=1}^{L}\xi(x,t)$.
The two-time auto-correlation function for $\eta$ in terms of fluctuations about mean $\delta \eta = \eta - \langle \eta\rangle$   is given by 
\begin{eqnarray}
\mathcal{C}_{\eta}(\tau) = \langle \delta \eta(t_1) \delta \eta(t_2)\rangle. 
\label{g_corr}
\end{eqnarray}
Our main goal is to understand how the correlation function for the local and total force fluctuations are related as a function of space and time. Note that a direct analytical computation of the two-time autocorrelation functions is a challenging problem. We employ the Wiener-Khinchin theorem, stating  that the Fourier transform of the correlation function of stationary process results in the power spectrum.  Taking Fourier transform of Eqs.~(\ref{l_corr}) and (\ref{g_corr}) with respect to $\tau$, we can get the power spectrum for local and total force fluctuations.

\begin{figure}[b]
  \centering
  \scalebox{0.60}{\includegraphics{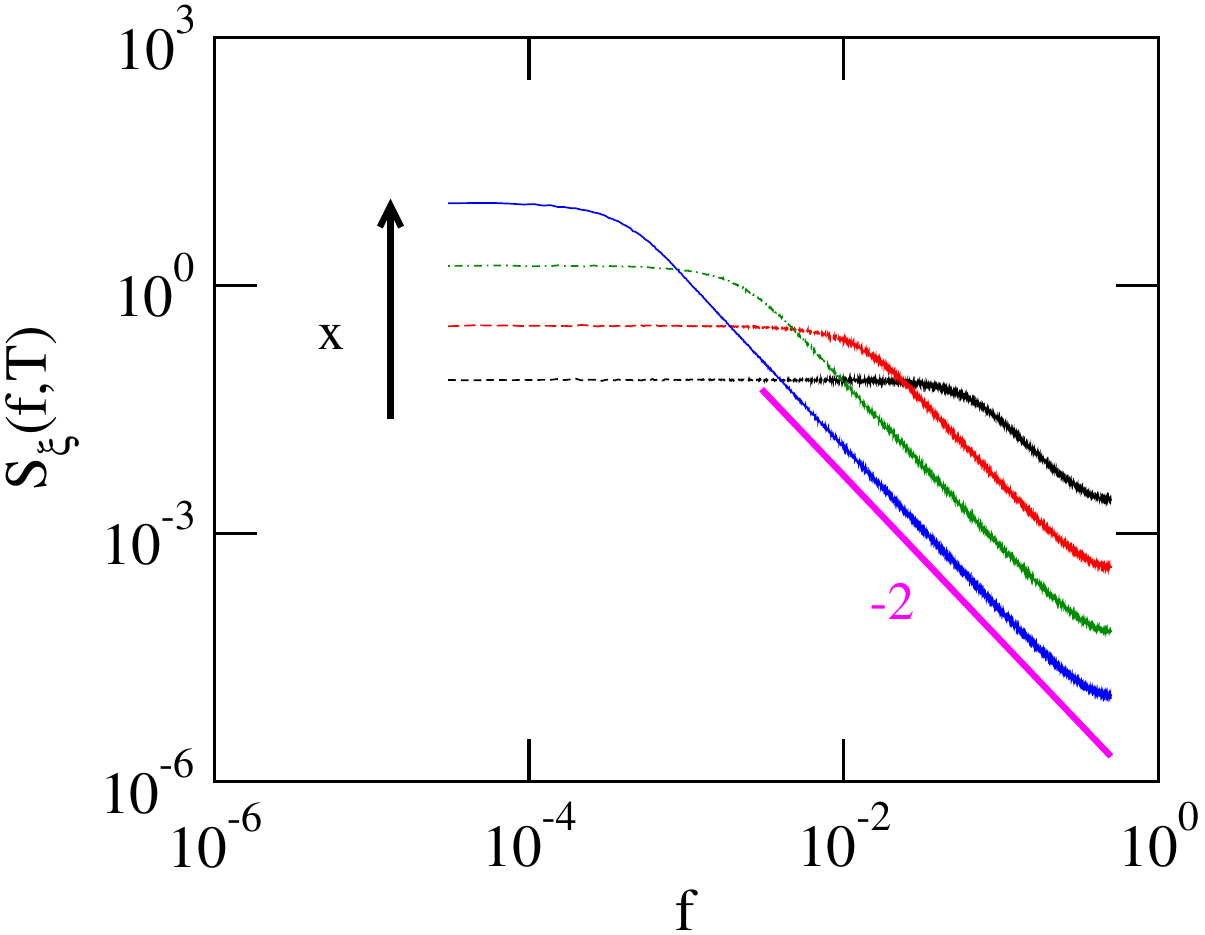}}
 \caption{The power spectra for the fluctuations in the local force $\xi(x,t)$ with different $x=$ $2^3, 2^5, 2^7$, and  $2^{9}$. The system size  is fixed  at $L = 2^{10}$. Here the cutoff time is $T \sim x^{\lambda}$. The signal length is $N=2^{16}$ and each curve is averaged over $10^4$ different realizations. }
  \label{fig_train_ps_xi_1}
\end{figure}

\section{Spectral properties}{\label{sec_3}}
In this Section, we show results for the power spectral density (PSD) properties for local and total force fluctuations. The  PSD is calculated as
 \begin{equation}
S(f) = \lim_{N\to \infty}\frac{1}{N}\langle |\hat{\xi}(f;N)|^2\rangle, 
\end{equation}
where $\hat{\xi}(f;N)$ is a Fourier-series transform of $\xi(t)$ on  time-interval  $t\in [0, N]$.  We numerically implement the standard fast Fourier transform methods to compute the power spectrum and use time series of length ranging from $2^{16}$ to $2^{18}$ after discarding the transient. We perform the ensemble average with $10^4$ independent realizations.  Our primary tool of analysis is the application of scaling methods that reveal an accurate behavior (as shown below).

\subsection{Fluctuations in local force: $\xi(x,t)$}
We compute spectral properties for the local force fluctuations  $\xi(x,t)$, where $x$ is the distance from the driving end. Our numerical results shown in Fig.~\ref{fig_train_ps_xi_1}  reveal that the power spectrum varies as $1/f^2$ for frequencies $1/T \ll f \ll 1/2$ with a cutoff. The cutoff time  $T$ is a function $x$. When the distance $x$ or equivalently the cutoff time is varied, the power spectra show explicit dependence on the time scale. The power spectrum is indeed a function of  $T$ for the entire range of frequencies.  Precisely, the power spectrum does not depend upon the frequency in the frequency regime $f\ll 1/T$. While the power value increases with increasing the cutoff time in this regime, the power decreases for $f\gg 1/T$.  Thus, we can mathematically write   
 \begin{equation}
S_{\xi}(f,T) = \begin{cases} AT^{2-\beta}, ~~~~~~{\rm for}~~f \ll 1/T, \\ A\frac{1}{f^2 }\frac{1}{T^{\beta}}, ~~~~~~{\rm for}~~1/T \ll f \ll 1/2, \end{cases}
\label{eq_z_ps}
\end{equation}
where $\beta$ is an exponent related to scaling as function of the cutoff time.

\begin{figure}[b]
  \centering
  \scalebox{0.60}{\includegraphics{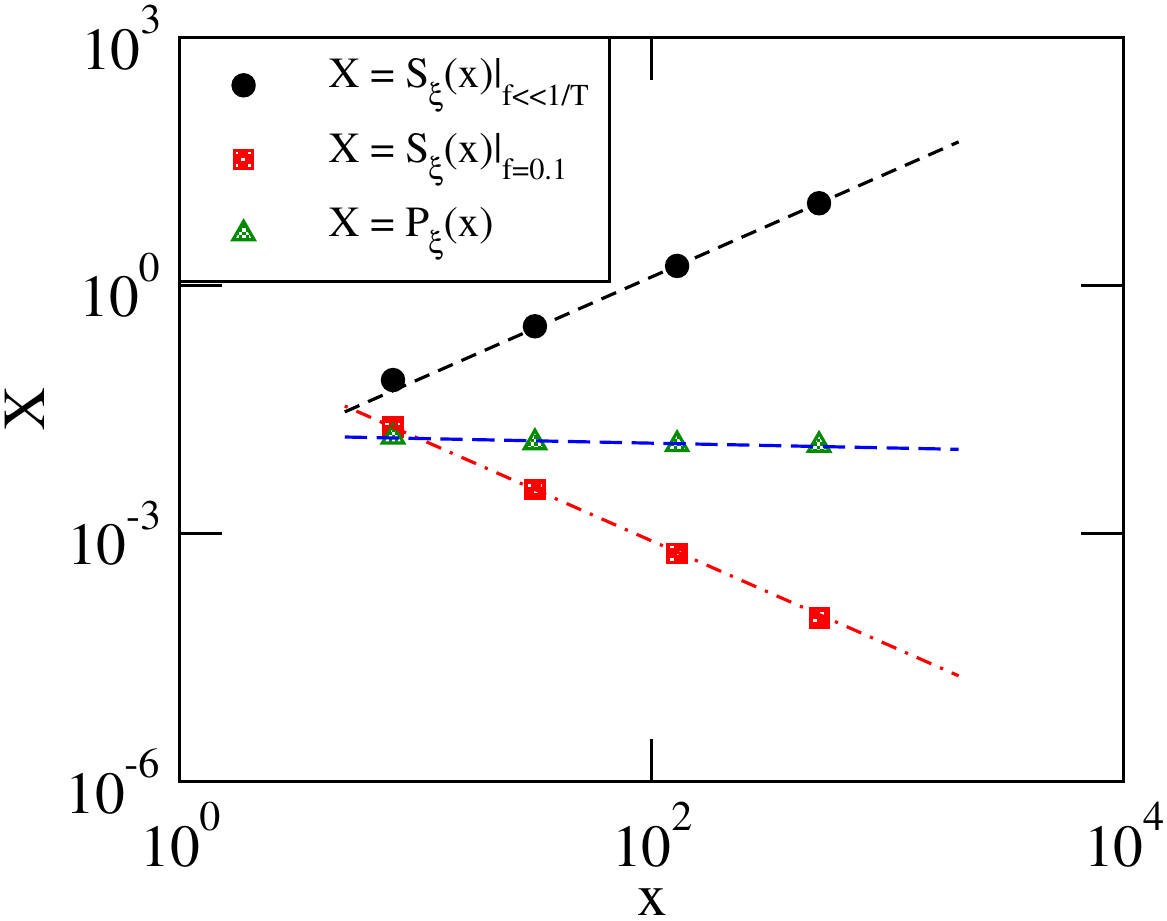}}
  \caption{The plots show the variation of power value in low frequency component $S_{\xi}(x)|_{f\ll1/T} \sim x^{(2-\beta) \lambda}$,  the power at fixed frequency  $S_{\xi}(x)|_{f\gg1/T} \sim x^{-\beta \lambda}$, and the total power $P_{\xi}(x)\sim x^{(1-\beta)\lambda}$ as a function of  $x$.   Using the best fit, the estimated slopes are $(2-\beta) \lambda = 1.253(8)$, $\beta \lambda = 1.255(5)$, and $(1-\beta)\lambda = 0.057(2)$. Thus, using the first two relations, the estimated value of the exponents are $\beta = 1.0008$ and $\lambda = 1.25$.}
  \label{fig_train_ps_xi_2}
\end{figure}

\begin{figure}[t]
  \centering
  \scalebox{0.60}{\includegraphics{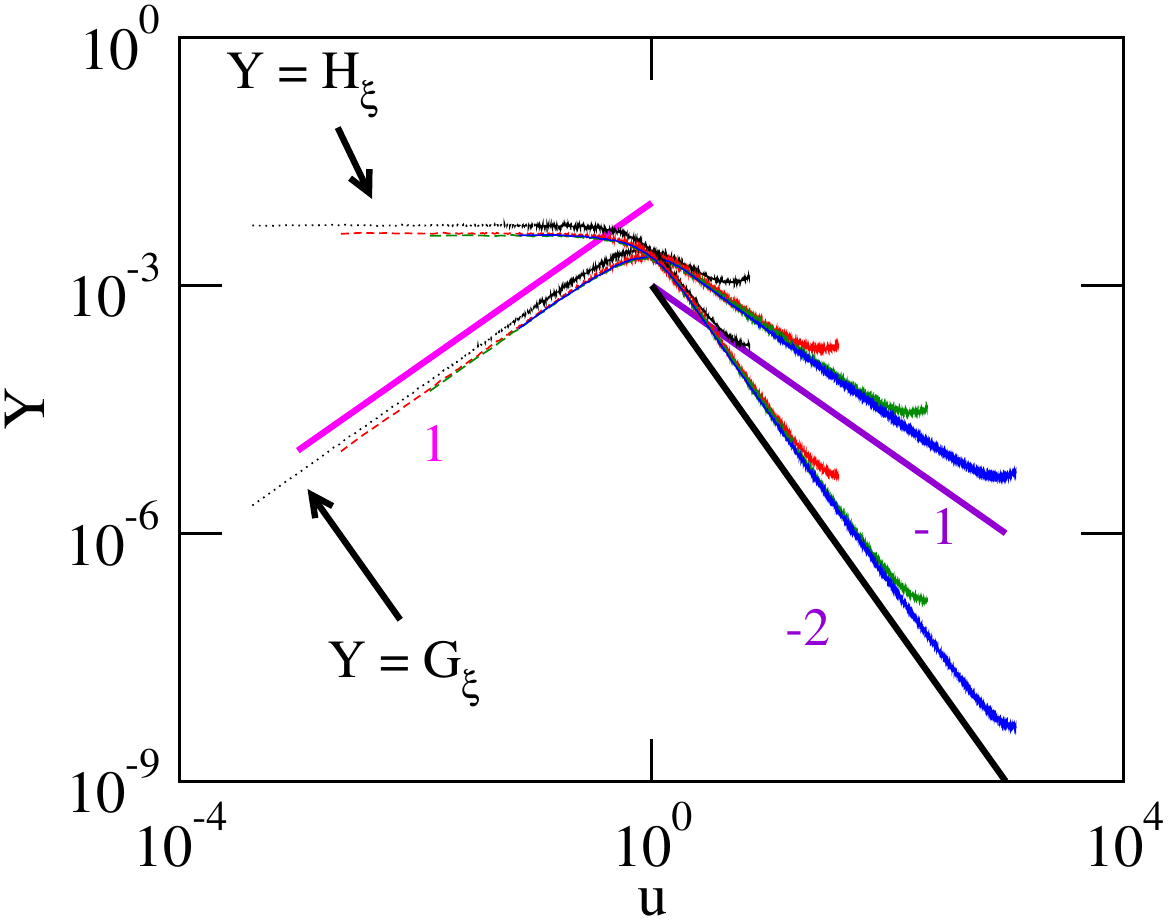}}
  \caption{The data collapse curves for the power spectra shown in Fig.~\ref{fig_train_ps_xi_1}. The scaling functions are: $G_{\xi}(u) \sim fS_{\xi}(f,T)$ and $H_{\xi}(u) \sim T^{-1}S_{\xi}(f,T)$ with $u = fT$.  }
  \label{fig_train_ps_xi_sf}
\end{figure}

Clearly, the power spectrum is a homogeneous function of the arguments, and one can apply the scaling methods~\cite{Yadav_2021} to re-express Eq.~(\ref{eq_z_ps}) as
 \begin{equation}
S_{\xi}(f,T) = A\frac{1}{f^{2-\beta}}G_{\xi}(u) = AT^{2-\beta} H_{\xi}(u),
\label{local_ps_scal}
\end{equation}
where $u = fT$. $G_{\xi}$ and $H_{\xi}$ are scaling functions. In order to estimate the exponent $\beta$, we compute the total power as a function of $T$
\begin{equation}
P_{\xi}(T) = \int df S_{\xi}(f,T) = \int df A\frac{1}{f^{2-\beta}}G_{\xi}(u) \sim T^{1-\beta}. \nonumber
\end{equation} 
It is noted that the cutoff time is related with the distance as $T \sim  x^{ \lambda}$, where $\lambda$ is an exponent. We numerically compute the total power as a function $x$ and note that this varies in a logarithmic manner [see Fig.~\ref{fig_train_ps_xi_2} for plot with $\triangle$ symbol]. This implies that the total power behaves as  $P_{\xi}(T)\sim T^{-\epsilon}$, with $\epsilon \to 0$. Thus, $\beta = 1+\epsilon$. 
Then, the asymptotic behavior of the scaling functions can be easily extracted
\begin{subequations}
\begin{align}
G_{\xi}(u) = \begin{cases} u, ~~~~~~{\rm for}~~u \ll 1, \\ 1/u, ~~~{\rm for}~~u\gg 1,\end{cases}
\label{eq_z_ga}
\end{align}
and
\begin{align}
H_{\xi}(u) = \begin{cases} 1, ~~~~~~{\rm for}~~u \ll 1, \\ 1/u^2, ~~{\rm for}~~u\gg 1.\end{cases}
\label{eq_z_h}
\end{align}
\end{subequations}
One can numerically estimate the value of $\lambda$, if the power at low frequency, $S_{\xi}(f\ll1/T) \sim T$, is plotted with $x$ [see Eq.~(\ref{eq_z_ps})]. 
Finally, we can determine the data collapse curves  with the help of $T$ and $\beta$. The scaling functions are shown in Fig.~\ref{fig_train_ps_xi_sf}, and these excellently satisfy Eqs.~(\ref{eq_z_ga}) and (\ref{eq_z_h}).

The power spectra of the local force fluctuations can also be understood in the following manner. As the underlying  exponentially decaying correlation function $C(\tau) \sim \exp(-|\tau/T|)$ with the relaxation or  the cutoff time $T$ implies that the corresponding spectrum is Lorentzian $S(f) \sim T/[1+(2\pi fT)^2]$. In the limiting conditions, the spectrum behaves as
\begin{equation}
S(f,T) \sim \begin{cases}T,~~~~~~~~~~~~~{\rm for}~~~ fT \ll 1,\\
1/(Tf^2),~~~~{\rm for}~~~fT\gg 1.\end{cases}
\label{eq_lorentz}
\end{equation}
Comparing Eqs.~(\ref{eq_z_ps}) and (\ref{eq_lorentz}), it is easy to recognize that the power spectra for the local force fluctuations are basically Lorentzian. Moreover, the $T$ is not a constant in the train model, it is a power-law function of distance from the driving end.

\subsection{Fluctuations in total force: $\eta(t)$}
 
\begin{figure}[t]
	\centering
	\scalebox{0.60}{\includegraphics{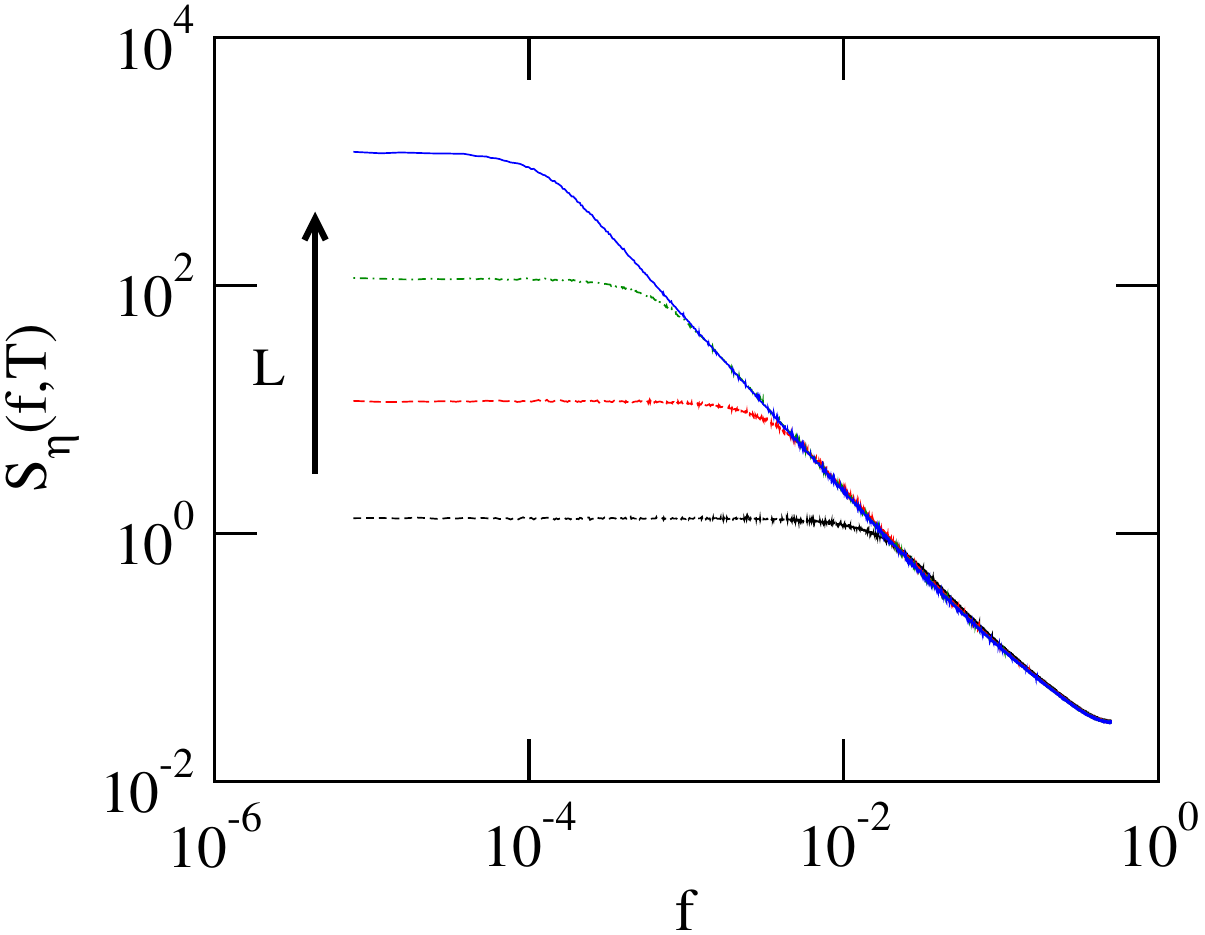}}
	\caption{The power spectra for $\eta(t)$ with different $L = 2^4, 2^6, 2^8,$ and $2^{10}$. Here $N = 2^{18}$. }
	\label{fig_train_ps_eta_01}
\end{figure}

\begin{figure}[t]
	\centering
	\scalebox{0.60}{\includegraphics{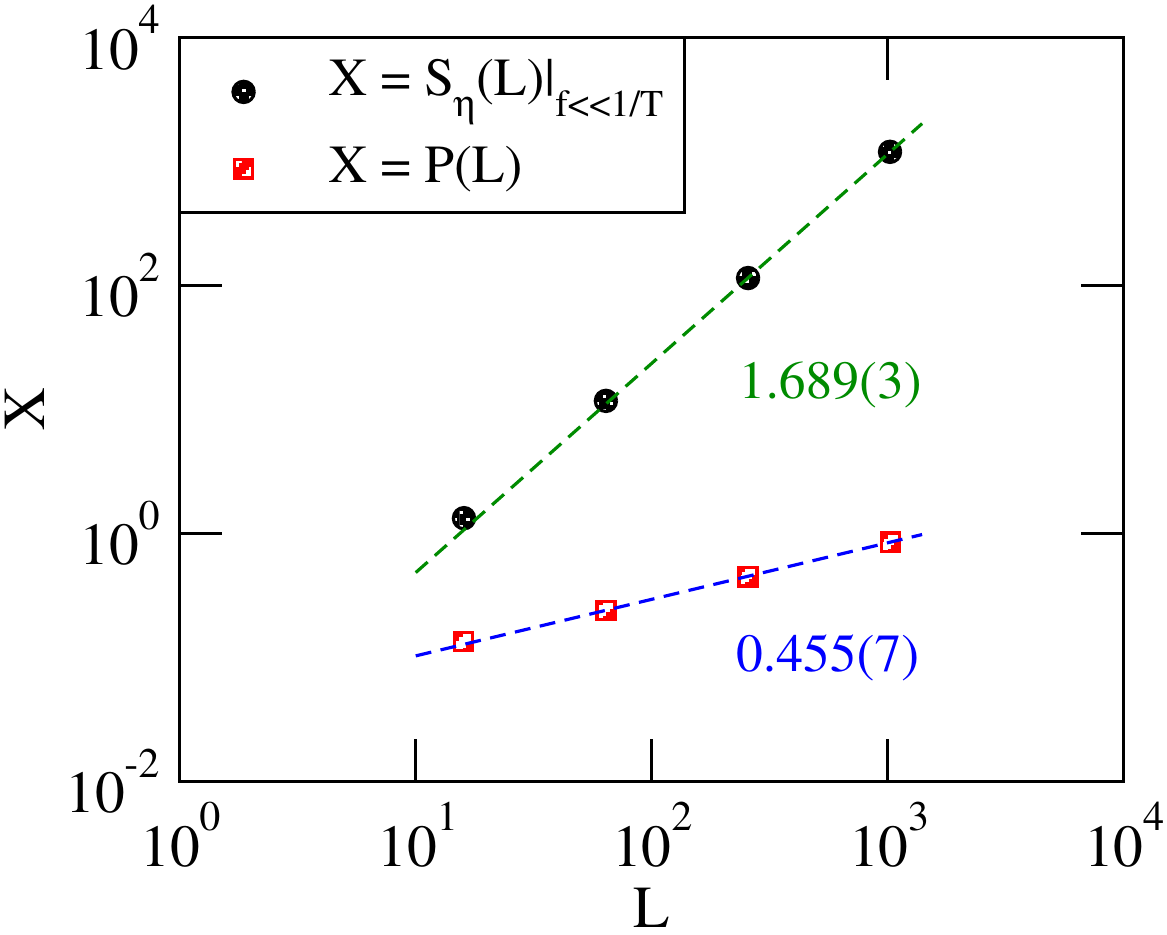}}
	\caption{The power at low frequency $S_{\eta}(L)|_{f\ll1/T} \sim L^{\alpha \lambda}$ and the total power $P(L)\sim L^{(\alpha-1)\lambda}$ as a function of the system size $L$.  The estimated value of the exponents are $\alpha = 1.37$ and $\lambda = 1.23$. }
	\label{fig_train_ps_eta_1}
\end{figure}

\begin{figure}[t]
	\centering
	\scalebox{0.60}{\includegraphics{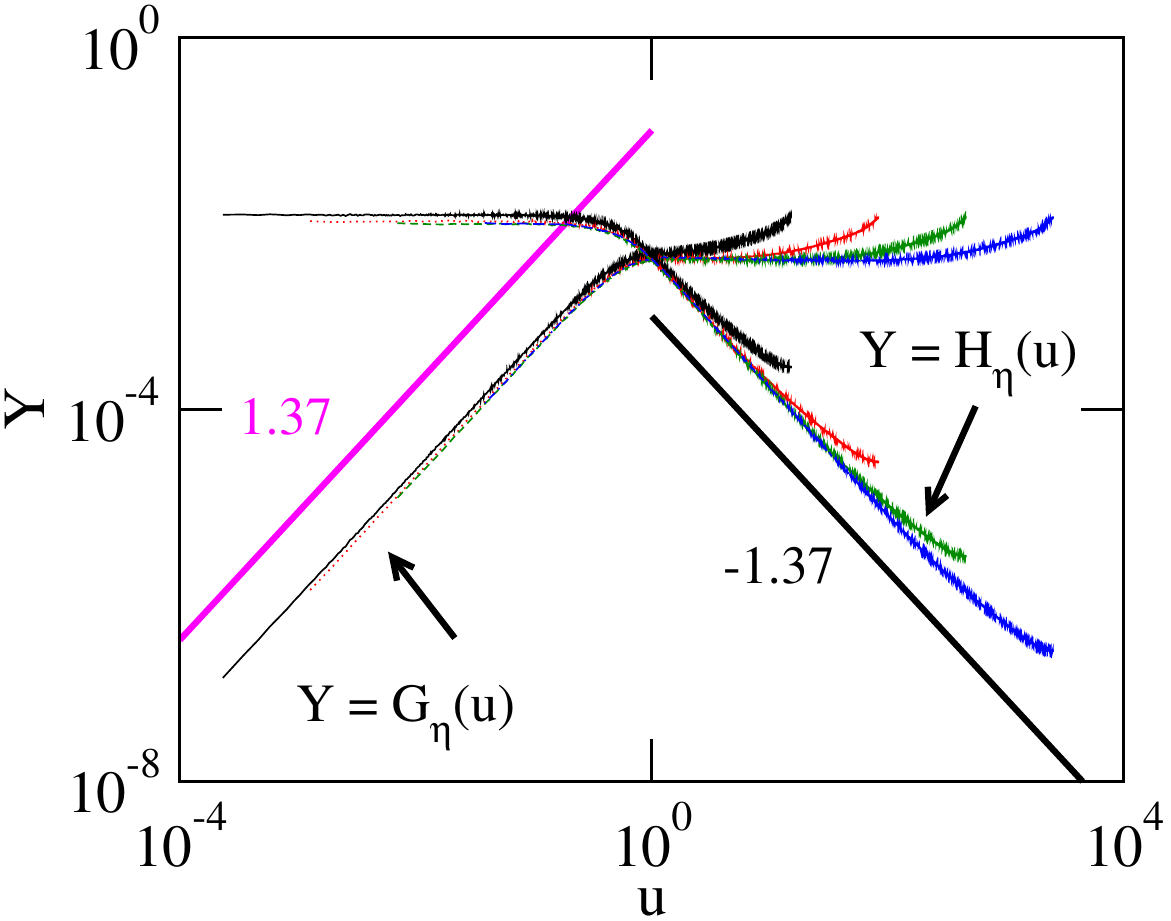}}
	\caption{The data collapse curves for the power spectra of total force fluctuations. Here, the scaling functions are: $G_{\eta}(u) = f^{\alpha}S_{\eta}(f,T)$ and $H_{\eta}(u) = T^{-\alpha}S_{\eta}(f,T)$ with $u = fT$.  }
	\label{fig_train_ps_eta_2}
\end{figure}

We also investigate the spectral content for the fluctuations in the total force $\eta(t)$.
We compute the power spectra for different system size $L$ and note $1/f^{\alpha}$ type behavior with a cutoff (see Fig.~\ref{fig_train_ps_eta_01}). The power does not depend on the cutoff time in the frequency regime $f\gg 1/T$, and the cutoff time varies with system size as $T \sim  L^{\lambda}$. The power spectrum depends upon the cutoff time only for frequencies $f\ll 1/T$.  Thus, we can write
 \begin{equation}
S_{\eta}(f,T) = \begin{cases} A'T^{\alpha}, ~~~{\rm for}~~f \ll 1/T, \\ A'\frac{1}{f^{\alpha}}, ~~~{\rm for}~~1/T \ll f \ll 1/2. \end{cases}
\end{equation}
The power spectrum is again a homogeneous function, and we can write using the scaling methods
 \begin{equation}
S_{\eta}(f,T) = A'\frac{1}{f^{\alpha}}G_{\eta}(u) = A'T^{\alpha} H_{\eta}(u),
\end{equation}
where $u = fT$. The scaling functions vary as
 \begin{subequations}
 \begin{align}
G_{\eta}(u) = \begin{cases} u^{\alpha}, ~~~~{\rm for}~~u \ll 1, \\ 1, ~~~~~~{\rm for}~~u\gg 1,\end{cases}
\label{eq_xi_g}
\end{align}
and
\begin{align}
H_{\eta}(u) = \begin{cases} 1, ~~~~~~~{\rm for}~~u \ll 1, \\ 1/u^{\alpha}, ~~{\rm for}~~u\gg 1.\end{cases}
\label{eq_xi_h}
\end{align}
\end{subequations}
The spectral exponent $\alpha$ can be alternatively estimated by plotting $S_{\eta}(f\ll 1/T)$ with $L$. Moreover, the total power as a function of the cutoff time $T$ varies as $P_{\eta}(T)  \sim T^{\alpha-1}$.
Figure~\ref{fig_train_ps_eta_1} shows the plot of total power as a function of $L$ or equivalently $T$.
The scaling functions are shown in Fig.~\ref{fig_train_ps_eta_2}. A good  data collapse indicates that the scaling functions satisfy Eqs.~(\ref{eq_xi_g}) and (\ref{eq_xi_h}) within the statistical error.
Since the total power varies as $P_{\eta}(T)  \sim T^{\alpha-1}$, the corresponding power spectrum is expected to behave as $S_{\eta}(f) \sim 1/f^{\alpha}$.

\section{The link for long-range space-time correlations}{\label{sec_4}}

If we assume the local fluctuations to be independent in space, one might expect that the power spectrum of the total force equals the sum of local spectra. We can compute this  using Eq.~(\ref{local_ps_scal})
\begin{eqnarray}
S_{\eta}(f) = \sum_xS_{\xi}[f,T(x)]\approx \int_{1}^{L} dx S_{\xi}(f,T)   = A \int \frac{dx}{ f} G_{\xi}(u), \nonumber
\end{eqnarray}
where $u = fT = fx^{\lambda}$. Then, $du = \lambda u x^{-1}dx$, with $x = T^{1/\lambda}$. Thus,
\begin{eqnarray}
S_{\eta}(f)  = \frac{A}{\lambda f^{1+1/\lambda}} \int du~u^{1/\lambda} \frac{G_{\xi}(u)}{u}
 \sim \frac{1}{f^{\alpha'}},
\end{eqnarray}
with $\alpha' = 1+1/\lambda$.

The numerical results suggest that $\alpha' \ne \alpha$. It implies that our assumption is not correct. In fact, the local fluctuations are not spatially independent but correlated. As the system displays SOC, we assume the spatial correlation to be decaying in a power-law manner 
$\mathcal{C}(x) \sim x^{-\gamma}$.
In turn, the power spectrum of the sum process can be expressed as 
\begin{eqnarray}
S_{\eta}(f) =  \int_{1}^{L} dx S_{\xi}(f,T)\mathcal{C}(x) \sim 1/f^{\alpha}.
\end{eqnarray}
Evaluating the integral, we note a scaling relation $\alpha = \alpha' -\gamma/\lambda = 1+(1-\gamma)/\lambda$.
Thus, the spatial correlation exponent is 
\begin{equation}
\gamma = [\alpha'-\alpha]\lambda = 1-(\alpha-1)\lambda.
\end{equation}
Clearly, if there were no spatial correlation, then $\alpha = \alpha'$.

\section{Oslo sandpile model}{\label{sec_5}}
The train model has been conjectured to belong to the universality class of Oslo sandpile model \cite{Christensen_1996}, which shows a map with the boundary driven interface depinning~\cite{Paczuski_1996}. Note that the Oslo sandpile model phenomenologically describes experiments on ricepiles~\cite{Frette_1996}. We here recall the definition of the Oslo sandpile model. Consider a one-dimensional lattice of linear extent $L$. Assign a discrete height variable to each site $h(x,t)$ and take the local slope to be $z(x,t) = h(x,t)-h(x+1,t) $. The right boundary is open $h(L+1)=0$, and the system is driven at left boundary by increasing height by one unit $h(1) = h(1) + 1$. If the local slope exceeds the threshold local slope $z^{c}(x)$, then the redistribution of height takes place as $h(x) = h(x) -1$ and $h(x+1) = h(x+1) +1$. Also, the threshold local slope is updated randomly to take a value of 1 or 2. This activity may further trigger the neighbor, and the neighbor may do the same. This event forms an avalanche. When the avalanche ends, the system is again driven.
 
Here, our interest is in the fluctuations at the external drive time scale. The quantities of interest may be the following: 
\begin{itemize}
	\item Local force: $\xi(x,t) = F(x,t) = z(x,t)-z^{c}(x,t)$. Total force $\eta(t) = \sum_x \xi(x,t)$. These observables are also relevant to interface fluctuations.
	\item Local slope: $\xi(x,t) = z(x,t)$.  Global slope: $\eta(x,t) = \sum_x \xi(x,t)$.
	
\end{itemize}

\begin{figure}[t]
  \centering
  \scalebox{0.57}{\includegraphics{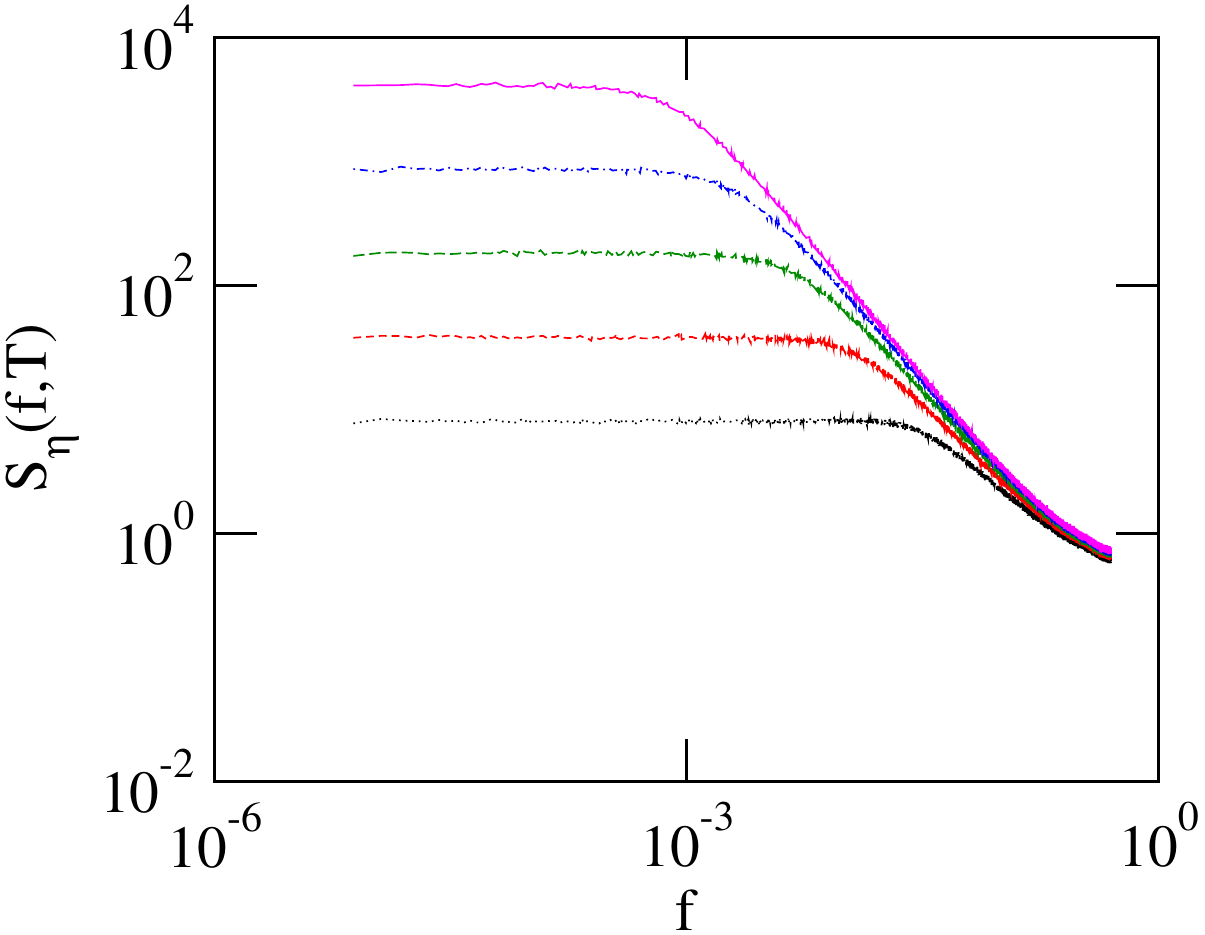}}
   \scalebox{0.57}{\includegraphics{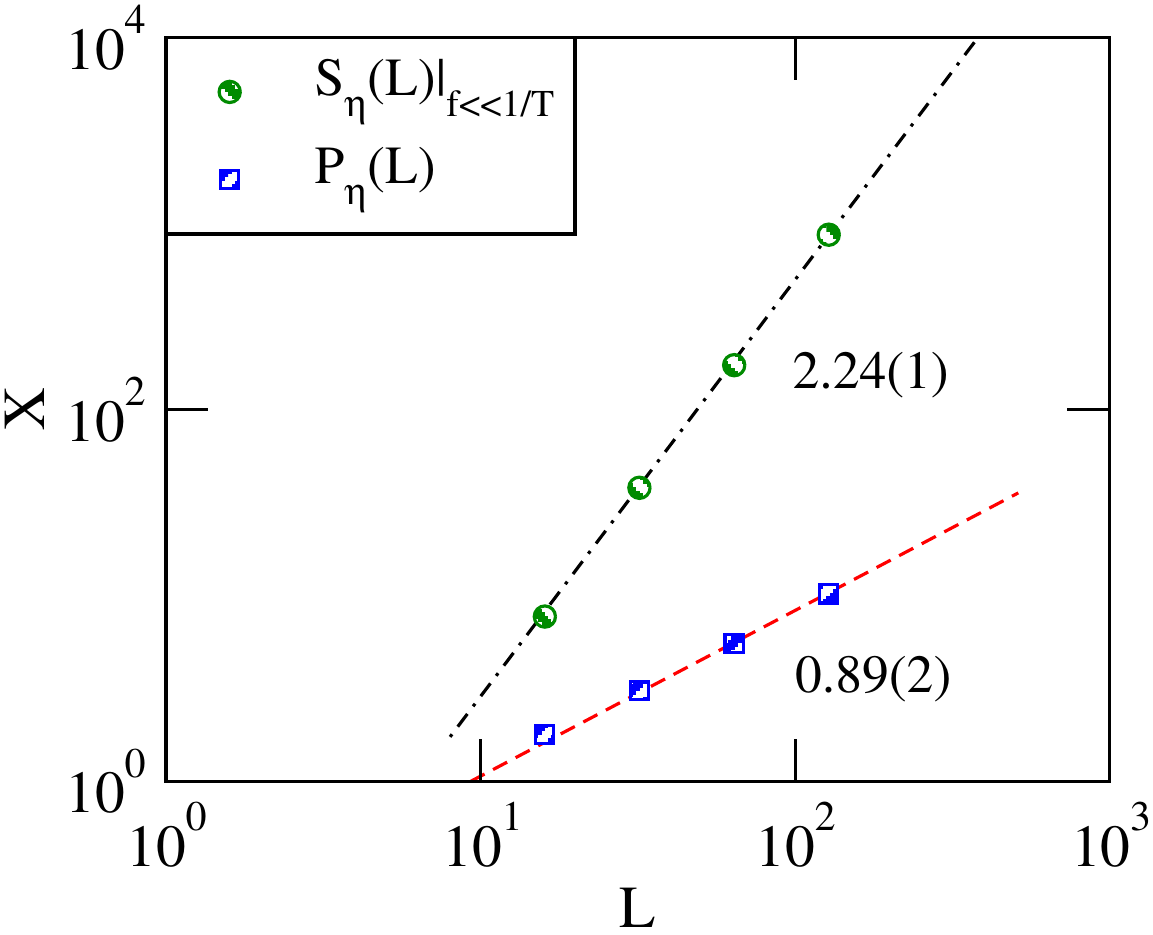}}  
  \caption{Top panel:  The power spectra for the fluctuations in the total force $\eta(t)$ with different system size $L$ $= 2^4, 2^5, 2^6, 2^7$, and  $2^{8}$. Here $T \sim L^{\lambda}$. The signal length is $N=2^{18}$ and the curves are averaged over $2\times 10^3$ different realizations. Bottom panel: The variation of the power value in low frequency component $S_{\eta}(L)|_{f\ll 1/T}\sim L^{(\alpha+\beta)\lambda}$ and the total power $P_{\eta}(L) \sim L^{(\alpha+\beta-1)\lambda}$ as a function of system size $L$. The estimated exponents are $\lambda = 1.35$ and $\alpha+\beta = 1.66$.}
  \label{fig_ps_eta_osp_0}
\end{figure}

\begin{figure}[t]
	\centering
	\scalebox{0.57}{\includegraphics{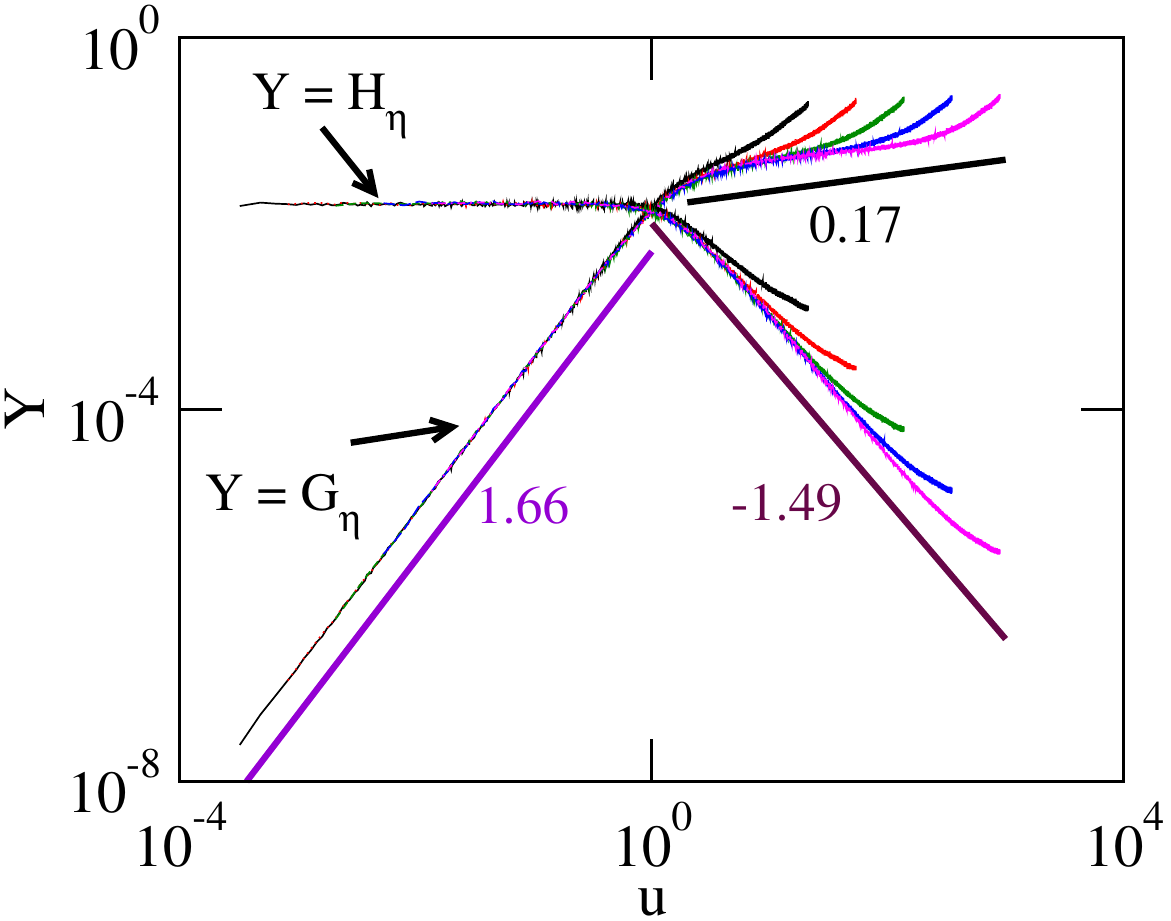}}
	\scalebox{0.57}{\includegraphics{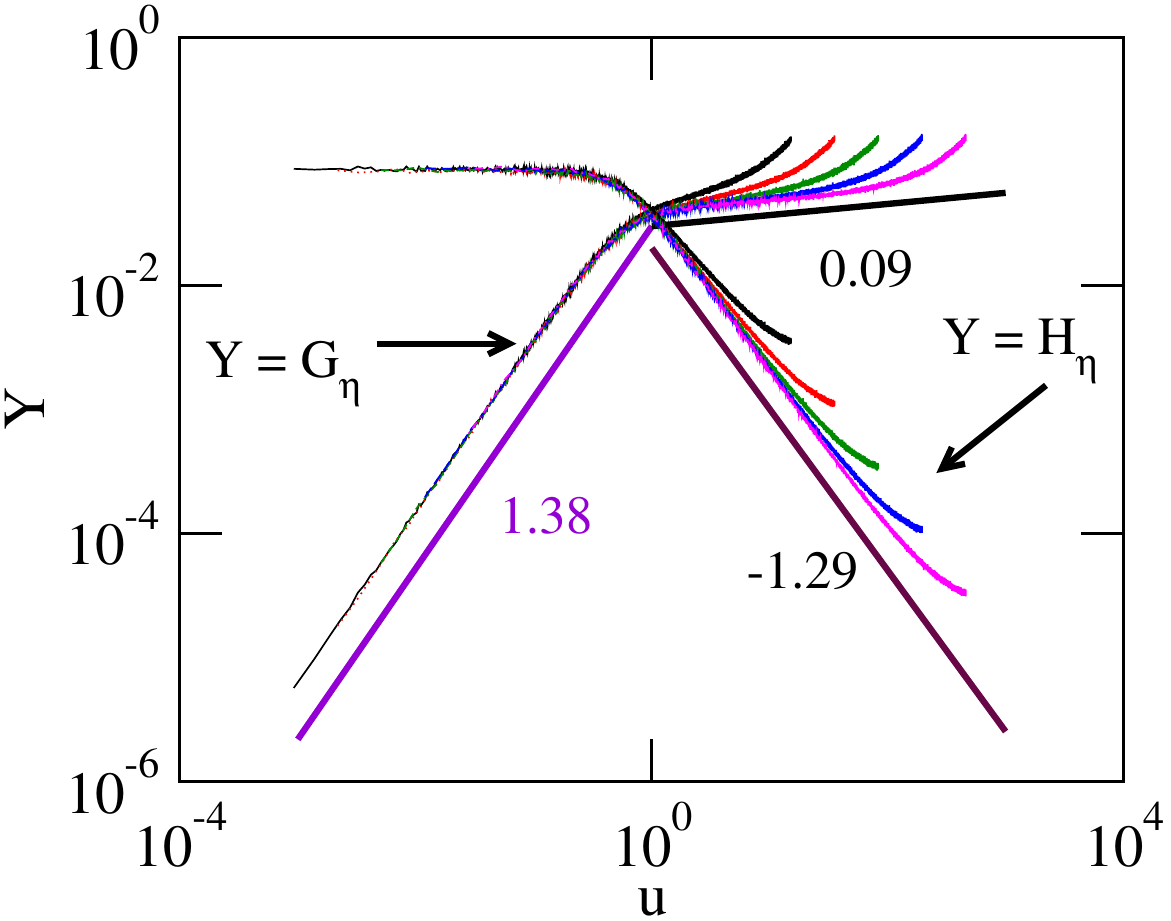}}
	\caption{The scaling functions for fluctuations in two quantities associated with Oslo sandpile model. Top panel: Total force. Bottom panel: Global slope. The corresponding spectral exponent $\alpha$ values are 1.49 and 1.29, respectively. }
	\label{fig_ps_eta_osp}
\end{figure}

We examine the power spectrum properties for the fluctuations in these quantities.
In both cases, the fluctuations in the local quantities $\xi(x,t)$ show $1/f^2$ behavior with an explicit cutoff time dependence as described by Eq.~(\ref{eq_z_ps}) to (\ref{eq_z_h}).   The results for the fluctuations in $\eta(t)$ (total force) show $1/f^{\alpha}$ behavior with a  marginal explicit cutoff time dependence, as can be seen in Fig.~\ref{fig_ps_eta_osp_0}. In this case, we note 
 \begin{equation}
S_{\eta}(f,T) = \begin{cases} A'T^{\alpha+\beta}, ~~~{\rm for}~~f \ll 1/T, \\ A'\frac{T^{\beta}}{f^{\alpha}},~~~ ~~~{\rm for}~~1/T \ll f \ll 1/2. \end{cases}
\end{equation}
The homogeneous feature of power spectrum implies
 \begin{equation}
S_{\eta}(f,T) = A'\frac{1}{f^{\alpha+\beta}}G_{\eta}(u) = A'T^{\alpha+\beta} H_{\eta}(u),
\end{equation}
where $u = fT$. The scaling functions are
\begin{subequations}
\begin{align}
G_{\eta}(u) = \begin{cases} u^{\alpha+\beta}, ~~~~~~{\rm for}~~u \ll 1, \\ u^{\beta}, ~~~~~~~~~{\rm for}~~u\gg 1,\end{cases}
\label{eq_z_g}
\end{align}
and
\begin{align}
H_{\eta}(u) = \begin{cases} 1, ~~~~~~~~{\rm for}~~u \ll 1, \\ 1/u^{\alpha}, ~~~~{\rm for}~~u\gg 1.\end{cases}
\label{eq_os_tf_sc}
\end{align}
\end{subequations}
We plot the scaling functions $G_{\eta}$ and $H_{\eta}$ in Fig.~\ref{fig_ps_eta_osp} (Top panel). Similarly, the scaling functions associated with global slope fluctuations are shown in  Fig.~\ref{fig_ps_eta_osp} (Bottom panel). Such a behavior is detectable if we employ the scaling methods. Since the power spectrum explicitly depends upon $T$ or equivalently $L$, we avoid considering an observable that explicitly depends on $L$.  While the spectral properties for the average slope have been examined \cite{Zhang_2000}, the analysis also requires examination of the effect due to explicit dependence on $L$.

\section{Conclusion}{\label{sec_6}}
To summarize, we have examined a class of sandpile models [train model and Oslo sandpile] that show the existence of non-trivial long-range space-time correlations. Our primary focus has been to understand a simple link between space and time correlations when the fluctuations are recorded at the external drive time scale.  Despite considerable efforts,  this important problem so far remained poorly understood.   To achieve this goal, we studied the power spectrum properties for both microscopic and macroscopic fluctuating observables. The fluctuations in the microscopic observables show $1/f^2$ type power spectra with an explicit cutoff time dependence.  In fact,  the microscopic variables show Lorentzian spectrum, where the cutoff time grows in a power--law  manner as function of distance from the driving end. On the other hand, the fluctuations in macroscopic observable  show non-trivial $1/f^{\alpha}$ type power spectrum.  
We note that except the train model, the power spectra for macroscopic observables show a marginal explicit cutoff time dependence.

It is interesting to mention that the scaling methods helped us to obtain precise spectral characteristics. It eventually leads to the identification of a simple link for the space-time correlations. To be specific, the link can be summarized in terms of a scaling relation among the spectral, cutoff time, and spatial correlation exponents.  Our systematic scaling analysis shown here offers a general approach to understand scale-invariant space-time correlations connecting microscopic and macroscopic fluctuations.  We expect that our systematic approach would be useful in many other contexts also.

\section*{ACKNOWLEDGMENT}
 NK would like to acknowledge financial support from the CUJ-UGC fellowship.
ACY acknowledges a grant ECR/2017/001702 funded by SERB, DST, Government of India.

\end{document}